\documentclass[twoside,a4paper,11pt]{sea9}
\usepackage{graphicx}
\usepackage{hyperref}
\def\aj{AJ}%
\def\araa{ARA\&A}%
\def\apj{ApJ}%
\def\apjl{ApJ}%
\def\apjs{ApJS}%
\def\aap{A\&A}%
\def\mnras{MNRAS}%
\def\procspie{Proc.~SPIE}%

\begin{document}
\pagenumbering{arabic}
\pagestyle{myheadings}
\thispagestyle{empty}
{\flushleft\includegraphics[width=\textwidth,bb=58 650 590 680]{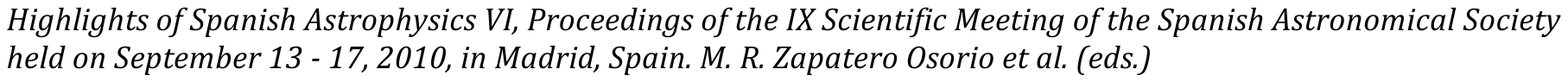}}
\vspace*{0.2cm}
\begin{flushleft}
{\bf {\LARGE
%
Detecting Galaxy Clusters in the DLS and CARS: a Bayesian Cluster Finder
%
}\\
\vspace*{1cm}
%
B. Ascaso$^{1}$,
D. Wittman$^{2}$, 
N. Ben\'itez$^{3}$,
and
the DLS collaboration.
%
}\\
\vspace*{0.5cm}
%
$^{1,2}$
Department of Physics, University of California, Davis, One Shields Ave., Davis, CA, 95616, USA\\
$^{3}$
Instituto de Astrof\'isica de Andaluc\'ia-CSIC, Glorieta de la Astronom\'ia s/n, 18008, Granada, Spain
%
\end{flushleft}
%
\markboth{
A Bayesian cluster finder
}{ 
%
Ascaso et al.
%
}
\thispagestyle{empty}
\vspace*{0.4cm}
\begin{minipage}[l]{0.09\textwidth}
\ 
\end{minipage}
\begin{minipage}[r]{0.9\textwidth}
\vspace{1cm}
\section*{Abstract}{\small
%
The detection of galaxy clusters in present and future surveys enables measuring mass-to-light ratios, clustering properties or galaxy cluster abundances and therefore, constraining cosmological parameters. We present a new technique for detecting galaxy clusters, which is based on the Matched Filter Algorithm from a Bayesian point of view. The method is able to determine the position, redshift and richness of the cluster through the maximization of a filter depending on galaxy luminosity, density and photometric redshift combined with a galaxy cluster prior.  We tested the algorithm through realistic mock galaxy catalogs, revealing that the detections are 100\% complete and 80\% pure for clusters up to z $<$1.2 and richer than $\Lambda \geq$  25 (Abell Richness $\geq$ 0). We applied the algorithm to the CFHTLS Archive Research Survey (CARS) data, recovering similar detections as previously published using the same data plus additional clusters that are very probably real.  We also applied this algorithm to the Deep Lens Survey (DLS), obtaining the first sample of optical-selected galaxy in this survey. The sample is complete up to redshift 0.7 and we detect more than 780 cluster candidates up to redshift 1.2. We conclude by discussing the differences between previous weak lensing detections in this survey and optical detections in both samples.
%
\normalsize}
\end{minipage}
%
%
%
\section{Introduction\label{intro}}

Clusters are cosmological probes for the formation and evolution of the Universe through measurement of their mass-to-light ratios or clustering properties. In addition, they are very useful astrophysical laboratories for the study of the properties of their galactic population. 

A number of methods to detect clusters have been developed based on the clusters' X-ray emission \cite{rosati02}, weak lensing \cite{tyson90,wittman01,wittman03} and Sunyaev-Zeldovich effect \cite{carlstrom02}.  Furthermore, a number of cluster detection methods based on optical data have provided a large dataset of clusters. These methods are based on modeling different properties of the clusters: geometric distribution of the galaxies \cite{ramella01,kim02,ramella02,lopes04,botzler04};  luminosity and density profiles \cite{postman96,postman02,gal03,eisenhardt08,milkeraitis10} and cluster red sequence, galaxy colors and brightest cluster galaxy (BCG) magnitudes \cite{gladders00,lopezcruz04,gladders05,koester07,wilson08}.

All of these different kind of methods have just put in evidence the need to have an unbiased cluster sample that covers both a wide range in mass and redshift in order to constrain the selection functions and the cosmological parameters of the Universe. A large number of studies of individual clusters at moderate and high redshift \cite{lopezcruz04,depropris04,ascaso08,ascaso09,harsono09}  have showed statistical trends even with wide dispersion for a number of cluster properties such as the color-magnitude relation (CMR), the blue fraction or the luminosity function. However, these clusters have often been selected to be at the high end of the mass function. The ideal goals are to determine the least unbiased sample of clusters, the highest completeness rates and the lowest cluster mass limit as possible.

\section{The Bayesian Cluster Finder}

The motivation of this work is to take advantage of all the characteristics of the present methods by modeling each cluster property. However, we still want to detect a cluster if one of these properties is not present, like for example the CMR in high redshift clusters. To accomplish this, we have designed a Bayesian cluster finder where each galaxy is assigned a Bayesian probability that the galaxy belongs to a cluster at a certain redshift. 

This probability can be decomposed into a likelihood which is based on a variation of the Matched Filter Algorithm \cite{postman96} including photo-z information, and a Bayesian prior, where we include previously known cluster properties (Ascaso et al. 2010b, ApJ, in prep.)

The likelihood models the probability that a galaxy with its position, photometric redshift, magnitude and morphological type belongs to a cluster at that position, with a given redshift and richness.  It is the product of the model probability for a cluster spatial profile  (we use a Plummer profile; \cite{postman96}), a luminosity function (we use a Schechter function, \cite{schechter76}) and a redshift probability distribution (either from a photometric redshift software  or from a Gaussian approximation). We show in Table  \ref{tab:all} the parameters  used in this work.

\begin{table*}[h]
      \caption{Likelihood parameters}
      \[
         \begin{array}{llllll}
\hline\noalign{\smallskip}
\multicolumn{2}{c}{\rm Density}&
\multicolumn{2}{c}{\rm Luminosity}&
\multicolumn{2}{c}{\rm Redshift}\\
\hline
\multicolumn{2}{c}{\rm Plummer}&
\multicolumn{2}{c}{\rm Schechter}&
\multicolumn{2}{c}{\rm Gaussian}\\
\hline\hline
\multicolumn{1}{c}{\rm r_c}&
\multicolumn{1}{c}{\rm r_{cut}}&
\multicolumn{1}{c}{\rm \alpha}&
\multicolumn{1}{c}{\rm M^*}&
\multicolumn{1}{c}{\rm \sigma}&
\multicolumn{1}{c}{\rm Centers}\\
\multicolumn{1}{c}{\rm (Kpc)}&
\multicolumn{1}{c}{\rm (Mpc)}&
\multicolumn{1}{c}{\rm }&
\multicolumn{1}{c}{\rm }&
\multicolumn{1}{c}{\rm }&
\multicolumn{1}{c}{\rm }\\
\hline
1.5 & 1.5 & -1.05 & -21.44 & 0.06(1+z) & 0.1-1.2\\
\hline
         \end{array}
      \]
\label{tab:all}
   \end{table*}

The prior enhances the probability that a cluster exists at a given position by including any a priori information about clusters. We  consider two main cluster properties to be included as prior information: The relation between the cluster CMR and its redshift and the BCG magnitude-redshift relation obtained from the MaxBCG sample of 13823 BCGs  \cite{koester07}.

We created redshift slices from 0.1 $\le z \le$1.2 in steps of 0.1. Then, for each redshift slice, we assigned each galaxy a probability. We fit a Gaussian to the overall probability for each redshift slice and use the peak and width of the Gaussian as the background and the dispersion respectively. Galaxies 3$\sigma$ above the background probability form the basis of cluster candidates. Then, we select the maximum probability as the initial center of the detection. We make radial density profiles and consider the limits of the cluster when the profile becomes constant. After that, we center iteratively the structure in the brightest galaxy within 1.5 Mpc. 

The output consists of a richness $\Lambda_{CL}$ (ie, the effective number of L$^*$ galaxies in the cluster, \cite{postman96}), the position of the cluster and the redshift slice which maximizes the probability. We also obtained a different estimation of the redshift by fitting a gaussian to the photometric redshift distribution of the galaxy population. 

Finally, we filtered those candidates by requiring that the difference between the two redshift estimates is smaller than $bin\times(1+z)$, i.e: the width of the bin multiplied by a factor depending on the redshift. We also merged two or more detections if they were closer than 1.5 Mpc and with estimated redshift difference smaller than 0.3.

\section{Simulations: Completeness and Purity}

We have performed simulations to test the reliability of the results. We simulated sets of clusters ranging  10 $\le \Lambda_{CL} \le$ 200 and 0.1 $\le z_{c} \le$ 1.2. The magnitudes were distributed by following a Schechter luminosity function with fixed parameters  $\alpha=-1.1$ and $M^*= -21$. The positions of the galaxies in the cluster were distributed according to a Plummer profile and the redshifts of the member galaxies were spread following a Gaussian function to mimic photometric redshift errors. Finally, we simulated the galaxy cluster colors by using the empirical cluster color distribution by \cite{baldry04}, combined with a shift to the expected color, obtained from synthetic spectral templates. 

These simulated clusters were embedded into a field galaxy distribution. The magnitudes, colors and photo-z distribution were taken from the original data and we redistributed the positions by following a Rayleigh-Levy galaxy pair distribution function. 

The results of the simulations show a completeness fraction always higher than 80\% for clusters up to redshift $<$ 1.2 and for every richness, while the purity is over $\approx$ 80\% for clusters  with $\Lambda_{CL} \ge 25$ up to redshift 1.2. 

We find that the purity decreases about 20\% for detections at z$<$0.6 and $\Lambda_{CL} \le 50$ if we do not include the prior information in our probability. However, the completeness stays approximately the same. Hence, the prior helps to discriminate the spurious detections that might be detected as clusters at low redshift and to increase purity.

In view of these results, to keep the completeness and purity rates over 80\%, we will consider real cluster detections those with $\Lambda_{CL} \ge 25$ up to redshift 1.2 with the prior turned on. 

\section{Application to Data}

We applied this algorithm to two different surveys, obtaining the following results.

The CFHTLS-Archive-Research Survey (CARS; \cite{erben09}) consists of catalogues obtained from the public archive images from the CFHTLS-Wide (37 square degrees).  The survey has five optical bands (ugriz), it is complete up to 24 in R band and it has photometric redshift available obtained from BPZ  \cite{benitez00}.

We detected galaxy clusters in this survey and compared the results with other surveys which use the CFHTLS data, such as \cite{olsen07} in CFHTLS-Deep (which overlaps one square degree with CARS) and  \cite{adami10} which used both Wide and Deep fields. We detected $\sim$ 90\% of the detections found in works  \cite{olsen07}  and \cite{adami10}, whereas \cite{adami10} and  \cite{olsen07} detect approx 70\%  and 80\% of our detections. In particular, we find a number of extra detections at z $\sim$ 0.9-1 that look probably real (Ascaso et al. 2010b). We plan to observe a sample of these candidates to confirm its existence. 

The Deep Lens Survey (DLS; \cite{wittman02})  is a very deep $BVRz$ imaging survey of five $2\times 2$ degree fields.  The depth is 27 mag arcsec$^{-2}$ in R and 26 mag arcsec$^{-2}$ in B,V and z. 

\begin{figure}
\center
\includegraphics[width=8.3cm,angle=0,clip=true]{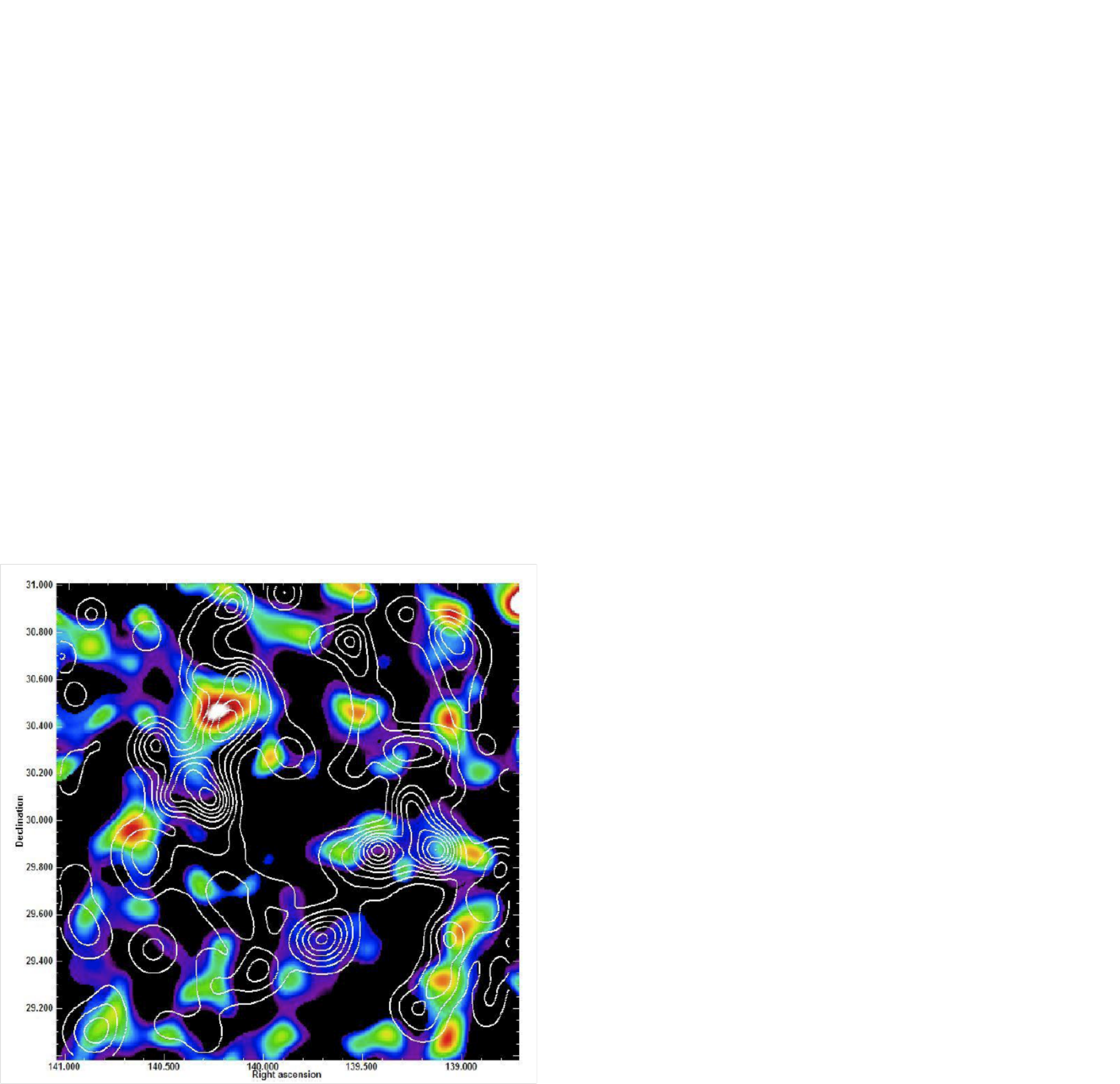} 
\caption{\label{fig:WL} DLS Field 2 (4 square degrees). The color map shows the weak lensing mass map, while the white contours refer to the optical galaxy cluster density map weighted by absolute luminosity and the weak lensing kernel. The most prominent structures are Abell 781 (upper left quadrant) at z=0.3, and a string of clusters at z=0.53 at lower right.}
\end{figure}

We detected more than 780 galaxy clusters in the whole DLS up to redshift 1.2, detecting 100\% of other optical detections from shallower galaxy cluster samples such as MaxBCG, \cite{koester07} up to z$<$0.6. We also detect 100\% of all the spectroscopically confirmed clusters \cite{wittman06} detected from Weak Lensing (WL). We are also able to map Large Scale Structure and compare the optical detections with the already available WL detections (Ascaso et al. 2010c, ApJ, in prep.). In Figure \ref{fig:WL}, we show the WL mass contours in color, with the the optical cluster detection weighted by the luminosity and the WL kernel in white. Large scale structures may be seen in both mass and weighted light. 

\section{Conclusions}

The Bayesian Cluster Finder is the first optical cluster detection technique that combines the Matched Filter Algorithm with a probability enhancement provided from cluster CMR and BCG magnitude if any.

We tested the method with simulations, which show 100\% completeness and $>$ 80\% purity up to z $<$1.2 for clusters richer than $\Lambda_{CL}>25$. We have successfully detected clusters in the DLS and CARS, providing good agreement with previous work and some additional detections at high redshift. This algorithm is ready to be applied to any survey. 
%
%
\small  
%
\section*{Acknowledgments}   
%
We are very grateful to the LOC and SOC for the excellent organization. BA and DW acknowledge the support of NASA grant NNG05GD32G.
%

%
\end{document}